\documentclass[12pt,a4paper]{article}
\usepackage{a4wide}
\usepackage{latexsym}
\usepackage{epsf}
\usepackage{amssymb}
\begin{document}

\begin{flushright}
\small
IFT-UAM/CSIC-05-06\\
{\bf hep-th/0501246}\\
January $31$st, $2005$
\normalsize
\end{flushright}

\begin{center}

\vspace{2cm}

{\Large {\bf A Note on Simple Applications}}
\vspace{.5cm}

{\Large {\bf  of the}}
\vspace{.5cm}

{\Large {\bf  Killing Spinor Identities}}

\vspace{2cm}

{\bf\large Jorge Bellor\'{\i}n}\footnote{E-mail: {\tt Jorge.Bellorin@uam.es}}
{\bf\large and Tom{\'a}s Ort\'{\i}n}\footnote{E-mail: {\tt Tomas.Ortin@cern.ch}}

\vspace{1cm}

{\it Instituto de F\'{\i}sica Te\'orica UAM/CSIC\\
  Facultad de Ciencias C-XVI,
  C.U. Cantoblanco,  E-28049-Madrid, Spain}\\

\vspace{3cm}


{\bf Abstract}

\end{center}

\begin{quotation}
  
  \small 
  
  We show how the Killing Spinor Identities (KSI) can be used to reduce the
  number of independent equations of motion that need to be checked explicitly
  to make sure that a supersymmetric configuration is a classical supergravity
  solution. We also show how the KSI can be used to compute BPS relations
  between masses and charges.

\end{quotation}

\newpage

\pagestyle{plain}


Supersymmetric solutions of supergravity theories play a very relevant role
today. As classical superstring backgrounds, they are used in the search for
phenomenologically viable superstring compactifications or, via the AdS/CFT
correspondence, they are used to study new states of SCFTs.

To find these solutions it is customary to introduce first an \textit{Ansatz}
that incorporates the relevant fields and symmetries into the Killing spinor
equations in order to constrain the form of the solution and make sure that
the required amount of supersymmetry will be preserved. Then one still has to
solve all the equations of motion, but this task is usually not too difficult
once the supersymmetry test is passed. It it, however, possible, to use the
Killing spinor equations in more efficient ways, as we are going to see.

For instance, recently, in Ref.~\cite{Lust:2004ig} it has been proven that,
for supersymmetric configurations of massive type~IIA supergravity, if the
equations of motion and Bianchi identities are satisfied for all the $p$-form
potentials and the dilaton, then the Einstein equations (and also the dilaton
equation) are also satisfied, under certain mild conditions. Similar results
had been obtained earlier in the context of minimal $d=5$ and $d=11$
supergravity in Refs.~\cite{Gauntlett:2002nw,Gauntlett:2002fz,Bandos:2003us}.
In this short note we are going to show that this result is a simple
consequence of the general \textit{Killing Spinor Identities} derived in
Ref.~\cite{Kallosh:1993wx}. These identities are relations between equations
of motion of the bosonic fields of supergravity theories and using them we can
show that the results of
Refs.~\cite{Lust:2004ig,Gauntlett:2002nw,Gauntlett:2002fz,Bandos:2003us} hold
in any theory of supergravity. These relations are the reason why
supersymmetric solutions depend on a very reduced number of independent
functions that solve simple equations. The advantage of this method is that it
is conceptually more clear and it does not require the computation of the
commutator of two supercovariant derivatives (the integrability conditions for
the Killing spinor equation), which is often algebraically quite involved.

The Killing Spinor Identities (KSI) of any supergravity theory with bosonic
and fermionic fields $\phi^{b},\phi^{f}$, and invariant under local
supersymmetry transformations
$\delta_{\epsilon}\phi^{b},\delta_{\epsilon}\phi^{f}$, can be derived as
follows: from the supersymmetry variation of the action of the theory, which
vanishes by hypothesis, we obtain the identity

\begin{equation}
\delta_{\epsilon}S = \int d^{d}x (S_{,b}\, \delta_{\epsilon}\phi^{b}   
+ S_{,f}\, \delta_{\epsilon}\phi^{f})=0\, .
\end{equation}

\noindent
Here $S_{,b (f)}$ are the first variations of the action with respect to the
bosonic (fermionic) fields, i.e.~their equations of motion.  Summation over
the indices $b,f$ is understood. Strictly speaking, the r.h.s.~of this formula
is a boundary term odd in fermion fields which we have assumed vanish on the
boundary. This is an acceptable assumption since we are going to set all the
fermionic fields to zero in the end.

Now we vary this equation w.r.t.~the fermionic fields and evaluate the
expression for vanishing fermionic fields, getting

\begin{equation}
\left\{S_{,bf_{2}}\, \delta_{\epsilon}\phi^{b}   
+S_{,b}\, (\delta_{\epsilon}\phi^{b})_{,f_{2}}
+S_{,f_{1}f_{2}}\, \delta_{\epsilon}\phi^{f_{1}}
+S_{,f_{1}}\, (\delta_{\epsilon}
\phi^{f_{1}})_{,f_{2}}\right\}_{\phi^{f}=0}=0\, .
\end{equation}

Since the bosonic equations of motion $S_{,b}$ and the supersymmetry
variations of the fermions $\delta_{\epsilon}\phi^{f}$ are necessarily
even in fermions

\begin{equation}
\left. S_{,bf_{2}}\right|_{\phi^{f}=0}=
\left. (\delta_{\epsilon}\phi^{f_{1}})_{,f_{2}}\right|_{\phi^{f}=0}=0\, ,  
\end{equation}

\noindent
and we are left with only two terms

\begin{equation}
\left\{S_{,b}\, (\delta_{\epsilon}\phi^{b})_{,f_{2}}
+S_{,f_{1}f_{2}}\, \delta_{\epsilon}\phi^{f_{1}}
\right\}_{\phi^{f}=0}=0\, .
\end{equation}

This expression is valid for any values of the bosonic fields $\phi^{b}$ and
supersymmetry parameters $\epsilon$, but it takes a most useful form when we
specialize it for supersymmetry parameters which are Killing spinors which we
denote by $\kappa$ and which satisfy, by definition, the Killing spinor
equation

\begin{equation}
  \left. \delta_{\kappa}\phi^{f}\right|_{\phi^{f}=0}=0\, .  
\end{equation}

Thus, supersymmetric (i.e.~admitting Killing spinors) bosonic configurations
satisfy the following \textit{Killing Spinor Identities} (KSI) found in
Ref.~\cite{Kallosh:1993wx} that relate their equations of motion

\begin{equation}
\left. S_{,b}\, (\delta_{\kappa}\phi^{b})_{,f}\right|_{\phi^{f}=0}=0\, .
\end{equation}

Of course, these equations are a particularly useful subset of the
supersymmetric gauge identities which relate all the equations of motion of a
locally supersymmetric theory, and their content is highly non-trivial even if
each term vanishes separately on-shell. This is the reason behind the
well-known fact that supersymmetric solutions are given in terms of a very
small number of functions that satisfy certain equations: each equation of
motion is a simple combination of the equations satisfied by those few
functions and that is how the equations of motion are related by the KSI, on-
or off-shell. For example, in simple $p$-brane solutions, all the equations of
motion are proportional to the Laplacian of a single function.

The KSI can be used, for instance, to reduce the number of independent
equations of motion that need to be solved explicitly\footnote{The contracted
  Bianchi identity $\nabla_{\mu}G^{\mu\nu}=0$ is used in General Relativity in
  a similar fashion: it implies
  $\sum_{\phi^{b}}\nabla_{\mu}T^{\mu\nu}(\phi^{b})=0$ and, given that
  $\nabla_{\mu}T^{\mu\nu}(\phi^{b})$ is always proportional to the equation of
  motion of the field $\phi^{b}$ (it only vanishes on-shell), we get a
  relation between the equations of motion of all the matter fields
  $\phi^{b}$. For a single minimally-coupled scalar field, for instance, if
  the Einstein equation is satisfied, we get
  $(\nabla^{2}\phi)(\nabla^{\nu}\phi)=0$ and, if $\nabla^{\nu}\phi\neq 0$ we
  get $\nabla^{2}\phi=0$, and if $\nabla^{\nu}\phi=0$ we get the same result.}
to make sure that a configuration satisfies them all. Let us consider a few
examples.

The action of the bosonic sector of $d=11$ supergravity
is\footnote{Our notation and conventions are those of
  Refs.~\cite{Janssen:1999sa} and \cite{Ortin:2004ms}.}

\begin{equation}
S= {\displaystyle\int}
 d^{11}x\
\sqrt{|g|}\ \left[R
-{\textstyle\frac{1}{2\cdot 4!}}G^{2}
-{\textstyle\frac{1}{(144)^{2}\sqrt{|g|}}} 
\epsilon G G C
\right]\, ,
\end{equation}

\noindent
and the supersymmetry variations of the bosonic fields are

\begin{equation}
\begin{array}{rcl}
\delta_{\epsilon} e{}^{a}{}_{\mu}
& = & 
-\frac{i}{2} \bar{\epsilon}\, \Gamma{}^{a}\, \psi_{\mu}\, , \\
& & \\
\delta_{\epsilon} C_{\mu\nu\rho}
& = & 
\frac{3}{2} \bar{\epsilon}\, \Gamma_{[\mu\nu}\, \psi_{\rho]}\, .
\end{array}
\end{equation}

\noindent
Defining

\begin{equation}
\begin{array}{rcccl}
E_{a}{}^{\mu}(e)\!\!\! & \equiv & \!\!\!
\frac{1}{\sqrt{|g|}}{\displaystyle\left.
\frac{\delta S}{\delta e^{a}{}_{\mu}}
\right|_{\psi=0}}\!\!\! & = & \!\!\!
-2
\left\{
G_{a}{}^{\mu} -{\textstyle\frac{1}{12}}
\left[G_{abcd}G^{\mu bcd}-{\textstyle\frac{1}{8}}e_{a}{}^{\mu}G^{2}
\right]
\right\} \, , \\
& & & & \\
E^{\mu\nu\rho}(C) \!\!\! & \equiv & \!\!\!
\frac{1}{\sqrt{|g|}}{\displaystyle
\left. 
\frac{\delta S}{\delta C_{\mu\nu\rho}}
\right|_{\psi=0}}\!\!\! & = & \!\!\!
{\textstyle\frac{1}{3!}}
\left[
\nabla_{\sigma}G^{\sigma\mu\nu\rho}
-{\textstyle\frac{1}{9\cdot 2^{7} \sqrt{|g|}}} 
\epsilon^{\mu\nu\rho\lambda_{1}\cdots\lambda_{4} 
\gamma_{1}\cdots \gamma_{4}}G_{\lambda_{1}\cdots\lambda_{4}}
G_{ \gamma_{1}\cdots \gamma_{4}}
\right] \, , \\
\end{array}
\end{equation}

\noindent
we immediately get the KSI of $d=11$ supergravity

\begin{equation}
\bar{\kappa}
\left[E_{a}{}^{\mu}(e)\gamma^{a}   
+3i E^{\mu ab}(C)
\gamma_{ab}\right]=0\, .
\end{equation}

\noindent 
If the equation of motion of the 3-form is satisfied (the Bianchi
identity is always assumed to be satisfied in this formalism), then, a
bosonic configuration always satisfies

\begin{equation}
\bar{\kappa} E_{a}{}^{\mu}(e)\gamma^{a}=0\, .
\end{equation}

\noindent
This is the equation obtained in
Ref.~\cite{Gauntlett:2002nw,Gauntlett:2002fz,Bandos:2003us,Lust:2004ig}
by computing the commutator of two supercovariant derivatives.  Now we
can follow the reasoning in
Refs.~\cite{Gauntlett:2002nw,Gauntlett:2002fz} to see under which
conditions this equation implies Einstein's $E_{a}{}^{\mu}(e)=0$.
Multiplying by $i\kappa$ on the right, we get

\begin{equation}
\label{eq:EV}
E_{a}{}^{\mu}V^{a}=0\, ,
\end{equation}

\noindent
where

\begin{equation}
V^{a}\equiv i\bar{\kappa}\gamma^{a}\kappa\, ,
\end{equation}

\noindent
is always a non-spacelike vector.  If we multiply by
$E_{b}{}^{\nu}(e)\gamma^{b}$ and symmetrize in the free indices we get

\begin{equation}
\label{eq:EE}
E_{a}{}^{\mu}(e)E_{b}{}^{\nu}(e)\eta^{ab}=0\, .  
\end{equation}

\noindent
If $V$ is spacelike, introducing a frame in which $e^{0}=V$, Eq.~(\ref{eq:EV})
implies that all the components $E_{0}{}^{\mu}(e)$ vanish\footnote{In
  Ref.~\cite{Lust:2004ig} this condition was imposed by hand. In this case, we
  see that it follows from Eq.~(\ref{eq:EV}). In the null case that we
  consider next, only part of this condition has to be imposed by hand.} and
Eqs.~(\ref{eq:EE}) can be seen as positive- or negative-definite scalar
products of vectors and one concludes that $E_{a}{}^{\mu}(e)=0$.

\noindent
If $V$ is null, we construct a frame 

\begin{equation}
ds^{2}=2e^{+}e^{-}-e^{i}e^{i}\, ,
\hspace{1cm}
i=1,\cdots,9\, .  
\end{equation}

\noindent
with $e^{+}=V$. Now Eq.~(\ref{eq:EV}) implies that all the components
$E_{-}{}^{\mu}(e)$ vanish and Eqs.~(\ref{eq:EE}) imply that
$E_{+}{}^{i}=E_{j}{}^{i}=0$. The only component of the Einstein
equation that one needs to impose independently is $E_{+}{}{}^{+}=0$.

Let us now consider the example directly studied in
Ref.~\cite{Lust:2004ig}: massive type~IIA supergravity. The action of
this theory is 

\begin{equation}
 \begin{array}{rcl}
S & = &
 {\displaystyle\int} d^{10}x\,
\sqrt{|g|} \biggl \{ e^{-2\phi}
\left[ R -4\left( \partial \phi \right)^{2}
+{\textstyle\frac{1}{2\cdot 3!}} H^{2}\right]
-\frac{1}{2}m^{2}
-{\textstyle\frac{1}{4}} G^{(2)\, 2}
-{\textstyle\frac{1}{2\cdot 4!}} G^{(4)\, 2}
\\
& & \\
& & 
-{\textstyle\frac{1}{144}} \frac{1}{\sqrt{|g|}}\
\epsilon\left[\partial C^{(3)}\partial C^{(3)}B 
+{\textstyle\frac{1}{2}}m\partial C^{(3)}BBB
+{\textstyle\frac{9}{80}}m^{2}BBBBB\right]
\biggr \}\, , \\
\end{array}
\end{equation}

\noindent
where the field strengths are given by

\begin{equation}
H = 3\partial B\, ,
\hspace{1cm}
G^{(2)}=2\partial C^{(1)}+m B\, ,
\hspace{1cm}
G^{(4)}  =  4\partial C^{(3)} -4HC^{(1)}+3m BB \, ,
\end{equation}

\noindent
and the supersymmetry transformation rules of the bosonic fields are 

\begin{equation}
\begin{array}{rcl}
\delta_{\epsilon} e{}^{a}{}_{\mu} & = & 
-i\bar{\epsilon}\Gamma^{a} \psi_{\mu}\, ,\\
& & \\
\delta_{\epsilon} B_{\mu\nu} & = & 
-2i\bar{\epsilon}\Gamma_{[\mu}
\Gamma_{11}\psi_{\nu]}\, , \\
& & \\
\delta_{\epsilon}\phi & = & 
-\frac{i}{2} \bar{\epsilon}\lambda\, ,\\
& & \\
\delta_{\epsilon} C^{(1)}{}_{\mu} & = & 
-e^{\phi}\bar{\epsilon} \Gamma_{11}
\left(\psi_{\mu} 
-\frac{1}{2}\Gamma_{\mu}\lambda \right)\, ,\\
& & \\
\delta_{\epsilon} C^{(3)}{}_{\mu\nu\rho}
& = & 
3 e^{\phi} \bar\epsilon \Gamma_{[\mu\nu}
\left(\psi_{\rho]} 
-\frac{1}{3!}\Gamma_{\rho]}\lambda \right) 
+3C^{(1)}{}_{[\mu}\delta_{\epsilon}B_{\nu\rho]}\, .\\
\end{array}
\end{equation}

\noindent 
The equations of motion of the different fields, using the same
notation as in the 11-dimensional case, are

\begin{equation}
  \begin{array}{rcl}
E_{\mu\nu}(e) & = & -2 e^{-2\phi}
\left\{
R_{\mu\nu} 
-2\nabla_{\mu}\nabla_{\nu}\phi 
+{\textstyle\frac{1}{4}} H_{\mu}{}^{\rho\sigma} H_{\nu\rho\sigma}
-{\textstyle\frac{1}{2}}e^{2\phi}\sum_{n=0,2,4} 
{\textstyle\frac{1}{(n-1)!}} T^{(n)}{}_{\mu\nu}
\right\}\\
& & \\
& & -\frac{1}{2}g_{\mu\nu}E(\phi)\, ,\\
& & \\
E(\phi) & = & 
-2e^{-2\phi}\left\{
R +4\left(\partial\phi\right)^{2}  -4\nabla^{2}\phi 
+{\textstyle\frac{1}{2\cdot 3!}}H^{2}
\right\} \\
& & \\
E^{\mu\nu}(B) & = & -\frac{1}{2}\{\nabla_{\rho}(e^{-2\phi}H^{\rho\mu\nu}) 
+m G^{(2)\, \mu\nu} +\frac{1}{2}G^{(4)\, \mu\nu\alpha\beta} G^{(2)}{}_{\alpha\beta}
 \\
& & \\
& & 
+\frac{1}{2\cdot (4!)^{2} \sqrt{|g|}}
\epsilon^{\mu\nu\alpha_{1}\cdots\alpha_{4}\beta_{1}\cdots\beta_{4}}
G^{(4)}{}_{\alpha_{1}\cdots\alpha_{4}} G^{(4)}{}_{\beta_{1}\cdots\beta_{4}}\}\\
& & \\
& & 
-3E^{\mu\nu\alpha}(C^{(3)}) C^{(1)}{}_{\alpha}\, ,
\\
& & \\
E^{\mu}(C^{(1)}) & = & \nabla_{\nu}G^{(2)\, \nu\mu}
+\frac{1}{3!}H_{\alpha_{1}\cdots\alpha_{3}}
G^{(4)\, \alpha_{1}\cdots\alpha_{3}\mu} \, , \\
& & \\
E^{\mu\nu\rho}(C^{(3)}) & = & \frac{1}{3!}
\{\nabla_{\sigma}G^{(4)\, \sigma\mu\nu\rho}
-\frac{1}{3!\cdot 4!  \sqrt{|g|}}
\epsilon^{\mu\nu\rho\alpha_{1}\cdots\alpha_{3}\beta_{1}\cdots\beta_{4}}
H_{\alpha_{1}\cdots\alpha_{3}} G^{(4)}{}_{\beta_{1}\cdots\beta_{4}} \}\, ,\\
  \end{array}
\end{equation}

\noindent 
where $T^{(n)}{}_{\mu\nu}$ are the energy-momentum tensors
of the RR fields:

\begin{equation}
T^{(n)}{}_{\mu\nu} = G^{(n)}{}_{\mu}{}^{\rho_{1}\cdots\rho_{n-1}}
G^{(n)}{}_{\nu\rho_{1}\cdots\rho_{n-1}}
-{\textstyle\frac{1}{2n}} g_{\mu\nu} G^{(n)\, 2}\, ,
\end{equation}

\noindent 
and, for $n=0$

\begin{equation}
T^{(0)}{}_{\mu\nu} = -{\textstyle\frac{1}{2}}m^{2} g_{\mu\nu}\, .
\end{equation}

The KSI of (massive) type~IIA supergravity associated to the
variations with respect to the gravitino and the dilatino take, then,
the form

\begin{equation}
  \begin{array}{rcl}
\bar{\kappa}\left\{E_{a}{}^{\mu}(e)\Gamma^{a} 
+2E^{a\mu}(B)\Gamma_{a}\Gamma_{11}
-ie^{\phi}E^{\mu}(C^{(1)})\Gamma_{11}
\right. & & \\
& & \\
\left. 
+3iE^{ab\mu}(C^{(3)})[e^{\phi}\Gamma_{ab}
-2iC^{(1)}{}_{a}\Gamma_{b}\Gamma_{11}] \right\} & = & 0\, ,\\
& & \\
\bar{\kappa}\{E(\phi) 
+ie^{\phi}E^{a}(C^{(1)})\Gamma_{11}\Gamma_{a}
-ie^{\phi}E^{abc}(C^{(3)})\Gamma_{abc}
 \} & = & 0\, .\\
\end{array}
\end{equation}

\noindent
The second equation tells us that, in presence of some unbroken
supersymmetries, if the equations of motion of the RR potentials are
satisfied, then the equation of motion of the dilaton is automatically solved.
If also the equation of motion of the NSNS 2-form is solved, then we get
$\bar{\kappa}E_{a}{}^{\mu}\Gamma^{a}=0$ as in the 11-dimensional case and,
following again the reasoning of Ref.~\cite{Gauntlett:2002fz} we arrive at the
same results.


By now, given that the Vielbein supersymmetry transformation rule
always has the same form, it should be clear that similar results are
going to hold in all supergravity theories.

For the sake of completeness we can also compute the KSI of type~IIB
supergravity. The equations of motion can be derived from the
non-self-dual (NSD) action of Ref.~\cite{Bergshoeff:1995sq}

\begin{equation}
\begin{array}{rcl}
S_{\rm NSD} \!\!& = &\!\!
{\displaystyle\int} d^{10}x\
\sqrt{|\jmath|}\ 
\biggl \{ e^{-2\varphi}
\left[ R(\jmath) -4\left( \partial \varphi \right)^{2}
+{\textstyle\frac{1}{2\cdot 3!}} {\cal H}^{2}\right]\\
& & \\
& & \!\!
+{\textstyle\frac{1}{2}} G^{(1)\, 2}
+{\textstyle\frac{1}{2\cdot 3!}} G^{(3)\, 2}
+{\textstyle\frac{1}{4\cdot 5!}} G^{(5)\, 2}
-{\textstyle\frac{1}{192}} \frac{1}{\sqrt{|\jmath|}}\
\epsilon\  \partial C^{(4)}\partial C^{(2)} {\cal B}
\biggr \}\, ,\\
\end{array}
\end{equation}

\noindent
where the field strengths are given by 

\begin{equation}
\mathcal{H} = 3\partial \mathcal{B}\, ,
\hspace{.5cm}
G^{(1)}=\partial C^{(0)}\, ,
\hspace{.5cm}
G^{(3)}  =  3\partial C^{(2)} -\mathcal{H} C^{(0)}\, ,
\hspace{.5cm}
G^{(5)}  =  5\partial C^{(4)} -10\mathcal{H} C^{(2)}\, .
\end{equation}

\noindent
The NSD action has to be supplemented, after variation, with the
self-duality of the 5-form field strength

\begin{equation}
G^{(5)}={}^{\star} G^{(5)}\, .  
\end{equation}

\noindent
The equations of motion that one derives from the NSD action are 

\begin{equation} 
  \begin{array}{rcl}
E_{\mu\nu}(e) & = & -2 e^{-2\varphi}
\left\{
R_{\mu\nu} 
-2\nabla_{\mu}\nabla_{\nu}\varphi 
+{\textstyle\frac{1}{4}} 
\mathcal{H}_{\mu}{}^{\rho\sigma} \mathcal{H}_{\nu\rho\sigma}
+{\textstyle\frac{1}{2}}e^{2\varphi}\sum_{n=1,3} 
{\textstyle\frac{1}{(n-1)!}} T^{(n)}{}_{\mu\nu}\right.
\\
& & \\
& & 
\left.+{\textstyle\frac{1}{4\cdot 4!}}e^{2\varphi}
T^{(5)}{}_{\mu\nu}
\right\}
-\frac{1}{2}\jmath_{\mu\nu}E(\varphi)\, ,\\
& & \\
E(\varphi) &=& -2 e^{-2\varphi}
\left\{ R + 4(\partial\varphi)^{2} -4 \nabla^{2} \varphi + 
\frac{1}{2\cdot 3!} {\cal H}^2 \right\}\, ,\\ 
& & \\
E^{\mu\nu} (\mathcal{B})
 &=& - \frac{1}{2} \left\{ \nabla_{\rho}
( e^{-2\varphi} \mathcal{H}^{\rho\mu\nu} ) -G^{(3)\, \mu\nu\alpha}G^{(1)}{}_{\alpha}
-\frac{1}{3!} G^{(5)+\, \mu\nu\alpha_{1}\alpha_{2}\alpha_{3}}
G^{(3)}{}_{\alpha_{1}\alpha_{2}\alpha_{3}}  \right\}\\
& & \\
& &  -C^{(0)}E^{\mu\nu}(C^{(2)})
-3!E^{\mu\nu\alpha\beta}(C^{(4)})C^{(2)}{}_{\alpha\beta}\, ,\\ 
& & \\
E(C^{(0)}) &=& - \left\{
\nabla_{\rho} G^{(1)\, \rho}
+\frac{1}{3!}G^{(3)\, \alpha\beta\gamma} \mathcal{H}_{\alpha\beta\gamma}\right\}\, , \\ 
& & \\
E^{\mu\nu}(C^{(2)}) &=& - \frac{1}{2} \{ \nabla_{\rho}
G^{(3)\, \rho\mu\nu} +\frac{1}{3!}
G^{(5)+\, \mu\nu\alpha_{1}\cdots \alpha_{3}}
\mathcal{H}_{\alpha_{1}\cdots \alpha_{3}} \}\, , \\  
& & \\
E^{\mu_{1}\cdots\mu_{4}}(C^{(4)}) &=& -\frac{1}{2\cdot 4!} \{ \nabla_{\rho} G^{(5)\,
\rho\mu_{1}\cdots\mu_{4}} -\frac{1}{(3!)^{2}\sqrt{|\jmath|}}
\epsilon^{\mu_{1}\cdots\mu_{4}\alpha_{1}\alpha_{2}\alpha_{3}\beta_{1}\beta_{2}\beta_{3}}
\mathcal{H}_{\alpha_{1}\alpha_{2}\alpha_{3}}G^{(3)}{}_{\beta_{1}\beta_{2}\beta_{3}}
\}\, .\\
\end{array}
\end{equation}

\noindent
The last equation is automatically satisfied once the self-duality of
$G^{(5)}$ is taken into account, and we will eliminate it from now on.
Taking the self-duality of $G^{(5)}$ into account the equation of
$\mathcal{B}$ also takes a simpler form:

\begin{equation}
  \begin{array}{rcl}
E^{\mu\nu} (\mathcal{B})
 &=& - \frac{1}{2} \left\{ \nabla_{\rho}
( e^{-2\varphi} \mathcal{H}^{\rho\mu\nu} ) -G^{(3)\, \mu\nu\alpha}G^{(1)}{}_{\alpha}
-\frac{1}{3!} G^{(5)\, \mu\nu\alpha_{1}\alpha_{2}\alpha_{3}}
G^{(3)}{}_{\alpha_{1}\alpha_{2}\alpha_{3}}  \right\}\\
& & \\
& &  
-C^{(0)}E^{\mu\nu}(C^{(2)})\, .\\
  \end{array}
\end{equation}

\noindent 
The supersymmetry variations of the bosonic fields are 

\begin{equation} 
  \begin{array}{rcl}
\delta_{\varepsilon} e_{\mu}{}^{a} & = & 
-i\bar{\varepsilon}\Gamma^{a} \zeta_{\mu}\, ,\\
& & \\
\delta_{\varepsilon}\varphi & = & 
-\frac{i}{2}\bar{\varepsilon}\chi\, ,\\
& & \\
\delta_{\varepsilon} {\cal B}_{\mu\nu} & = &
-2i\bar{\varepsilon} \sigma^{3} \Gamma_{[\mu} 
\zeta_{\nu]}\, , \\
& & \\
\delta_{\varepsilon} C^{(0)} &=& 
\frac{1}{2} e^{-\varphi} \bar{\varepsilon} \sigma^{2}  \chi\, , \\  
& & \\
\delta_{\varepsilon} C^{(2)}{}_{\mu\nu} & = & 
2i e^{-\varphi} \bar{\varepsilon} \sigma^{1} \Gamma_{[\mu} \left( \zeta_{\nu]} -\frac{1}{4} 
\Gamma_{\nu]} \chi \right) + C^{(0)} \delta_{\varepsilon} \mathcal{B}_{\mu\nu}\, , \\  
& & \\
\delta_{\varepsilon} C^{(4)}{}_{\mu\nu\rho\sigma}  & = & 
-4 e^{-\varphi} \bar{\varepsilon} \sigma^{2} \Gamma_{[\mu\nu\rho} \left( \zeta_{\sigma]} 
-\frac{1}{8}\Gamma_{\sigma]} \chi \right) 
+6 C^{(2)}{}_{[\mu\nu} \delta_{\varepsilon} \mathcal{B}_{\rho\sigma]}\, ,\\
\end{array}
\end{equation} 

\noindent
and the KSI of type~IIB supergravity are given by

\begin{equation} 
  \begin{array}{rcl}
{\bar \kappa} \{E_{a}{}^{\mu}(e)\Gamma^{a} +E^{a\mu}(\mathcal{B}) 
\sigma^{3} \Gamma_{a} -2E^{a\mu}(C^{(2)})
[ e^{-\varphi} \sigma^{1} -C^{(0)} 
\sigma^{3} ]\Gamma_{a} \} &= & 0\, ,\\
& & \\
{\bar \kappa} \{ 
E(\varphi) + i E(C^{(0)}) 
e^{-\varphi} \sigma^{2} + E^{ab}(C^{(2)}) e^{-\varphi} 
\sigma^{1} \Gamma_{ab} \} & = & 0\, . \\ 
\end{array}
\end{equation}

\noindent
If the equation of $C^{(2)}$ is satisfied, those of the two scalars
$\varphi,C^{(0)}$ are automatically satisfied. Further, if the
equation of $\mathcal{B}$ is satisfied, we arrive again at
$\bar{\kappa}E_{a}{}^{\mu}(e)\Gamma^{a}=0$.


Another use (the one originally proposed in
Ref.~\cite{Kallosh:1993wx}) is to constrain the form of corrections
(due to quantum effects or to the presence of external sources) to
supersymmetric solutions. The main assumption here is that the
supersymmetry transformation rules themselves do not get any
corrections. Under these conditions, if the bosonic fields satisfy now
the equations

\begin{equation}
S_{, b}= J_{b}\, ,  
\end{equation}

\noindent
then the sources $J_{b}$ must satisfy

\begin{equation}
\left. J_{b}\, (\delta_{\kappa}\phi^{b})_{,f}\right|_{\phi^{f}=0}=0\, .
\end{equation}

\noindent
Since the integration of the sources gives the charges of the object
that generates the fields of the solution, the KSI identities give BPS
relations between those charges. Observe that this method does not
allow for magnetic sources or charges, since the Bianchi identities
are assumed to hold from the beginning, although perhaps it might be
generalized to overcome this problem.

Let us consider a simple example: $N=2,d=4$ ungauged supergravity. The
action for the bosonic fields $g_{\mu\nu}, A_{\mu}$ is

\begin{equation}
S = \int d^{4}x \sqrt{|g|}\, [R-{\textstyle\frac{1}{4}}F^{2}]\, ,
\hspace{1cm}
F=2\partial A\, ,
\end{equation}

\noindent
and the supersymmetry variations of the bosonic fields are

\begin{equation}
  \begin{array}{rcl}
\delta_{\epsilon} e^{a}{}_{\mu} & = & -i\bar{\epsilon}\gamma^{a}\psi_{\mu} +c.c.\, , \\
& & \\ 
\delta_{\epsilon} A_{\mu} & = & -2i\bar{\epsilon}\psi_{\mu} +c.c.\, .\\
  \end{array}
\end{equation}

\noindent 
The equations of motion are 

\begin{equation} 
  \begin{array}{rcl}
E_{a}{}^{\mu}(e) &=& -2 \{ G_{a}{}^{\mu} 
-\frac{1}{2} \left[F_{ab} F^{\mu b} - \frac{1}{4} e_{a}{}^{\mu} F^{2}\right] 
\}\, , \\ 
& & \\
E^{\mu}(A) & = &  \nabla_{\alpha} F^{\alpha\mu}\, , \\
\end{array}
\end{equation}

\noindent
and the KSI  are given by

\begin{equation} 
\bar{\kappa}\{ E_{a}{}^{\mu} (e) 
\gamma^{a} +2E^{\mu}(A) \} = 0\, . 
\end{equation}

\noindent
These equations lead to relations between sources as those found in
Refs.~\cite{Tod:1983pm} and \cite{Caldarelli:2003wh} in which off-shell
configurations of $N=2,d=4$ ungauged and gauged supergravity were considered.

Observe that, according to the standard argument, in the timelike case, these
equations tell us that one only has to solve the Maxwell equations and Bianchi
identities for the vector field strength in order to have a solution of the
full set of equations of motion, and these equations reduce to just two
equations for two real functions (combined into a complex function thanks to
electric-magnetic duality). The same argument goes through in the gauged case,
studied in Refs.~\cite{Caldarelli:2003pb,Cacciatori:2004rt}, where it can be
seen that there are only two equations for two real functions
becasue the extra real function and the equation that it satisfies can be
deduced from the other two.

Defining sources for the fields $E_{a}{}^{\mu}(e)\equiv 2
T_{a}{}^{\mu}$ and $E^{\mu}(A)=J^{\mu}$ and multiplying the KSI by
$i\kappa$ from the right gives

\begin{equation} 
T_{a}{}^{\mu} (e) 
V^{a} +a J^{\mu}(A)  = 0\, , 
\end{equation}

\noindent
where we have defined the real bilinears

\begin{equation}
V^{a}=i\bar{\kappa}\gamma^{a}\kappa\, ,
\hspace{1cm}
a = i\bar{\kappa}\kappa\, .  
\end{equation}

\noindent
Let us now make assume that

\begin{enumerate}
\item Our supersymmetric configuration satisfies the condition that
  all the components $E_{a}{}^{\underline{0}}(e)$, $a\neq 0$ vanish
  (which is valid for the kind of static configuration that we have in
  mind in this simple example). Then, taking $\mu=0$ in the above
  equation, we get

\begin{equation} 
T_{0}{}^{\underline{0}} (e) 
V^{0} +a J^{\underline{0}}(A)  = 0\, , 
\end{equation}

\item The Killing spinor satisfies a projection condition of the form

  \begin{equation}
  (1\pm \gamma^{0})\kappa =0\, .  
  \end{equation}

\end{enumerate}

Then, $V^{0}=\mp a$ and we get a relation between gravitational and electric sources  

\begin{equation} 
T_{0}{}^{\underline{0}} (e) 
\mp J^{\underline{0}}(A) = 0\, , 
\end{equation}

\noindent
that will give $M=|Q|$ upon integration.

Clearly, similar arguments a and use of projectors lead to the
relation between mass and charge of the M2-brane in 11-dimensional
supergravity.


\vspace{1cm}

\textbf{Acknowledgements.} The authors are indebted to E.K.~Mata and
M.M.~Fern\'andez for their support. This work has been supported in
part by the Spanish grant BFM2003-01090.  

\vspace{1cm}



\begin{thebibliography}{99}


\bibitem{Lust:2004ig}
D.~L\"ust and D.~Tsimpis,
arXiv:hep-th/0412250.

\bibitem{Gauntlett:2002nw}
J.~P.~Gauntlett, J.~B.~Gutowski, C.~M.~Hull, S.~Pakis and H.~S.~Reall,
Class.\ Quant.\ Grav.\  {\bf 20} (2003) 4587
[arXiv:hep-th/0209114].

\bibitem{Gauntlett:2002fz}
J.~P.~Gauntlett and S.~Pakis,
JHEP {\bf 0304} (2003) 039
[arXiv:hep-th/0212008].

\bibitem{Bandos:2003us}
I.~A.~Bandos, J.~A.~de Azc\'arraga, J.~M.~Izquierdo, M.~Pic\'on and O.~Varela,
Phys.\ Rev.\ D {\bf 69} (2004) 105010
[arXiv:hep-th/0312266].

\bibitem{Kallosh:1993wx}
R.~Kallosh and T.~Ort\'{\i}n,
arXiv:hep-th/9306085.

\bibitem{Janssen:1999sa}
B.~Janssen, P.~Meessen and T.~Ort\'{\i}n,
Phys.\ Lett.\ B {\bf 453} (1999) 229
[arXiv:hep-th/9901078].

\bibitem{Ortin:2004ms}
T.~Ort\'{\i}n,
\textit{Gravity and strings},
Cambridge University Press (2004).

\bibitem{Bergshoeff:1995sq}
E.~Bergshoeff, H.~J.~Boonstra and T.~Ort\'{\i}n,
Phys.\ Rev.\ D {\bf 53} (1996) 7206
[arXiv:hep-th/9508091].

\bibitem{Tod:1983pm}
K.~P.~Tod,
Phys.\ Lett.\ B {\bf 121} (1983) 241.

\bibitem{Caldarelli:2003wh}
M.~M.~Caldarelli and D.~Klemm,
Class.\ Quant.\ Grav.\  {\bf 21} (2004) L17
[arXiv:hep-th/0310081].

\bibitem{Caldarelli:2003pb}
M.~M.~Caldarelli and D.~Klemm,
JHEP {\bf 0309} (2003) 019
[arXiv:hep-th/0307022].

\bibitem{Cacciatori:2004rt}
S.~L.~Cacciatori, M.~M.~Caldarelli, D.~Klemm and D.~S.~Mansi,
JHEP {\bf 0407} (2004) 061
[arXiv:hep-th/0406238].





                          
\end{thebibliography}
\end{document}